\documentstyle[12pt]{article}
\begin{document}
\title{ {\small \tt to appear in "The Physics of Sliding Friction", Ed.
B.N.J. Persson (Kluwer Academic Publishers), 1995 } \\
\vspace{1cm}
Sliding Friction in the Frenkel-Kontorova Model }

\author{ E. Granato $^{\dag}$ \\
I.N.P.E., S\~ao Jos\'e dos Campos,  S\~ao Paulo, Brazil\\
and I.C.T.P., Trieste, Italy\\
M.R.  Baldan\\
I.N.P.E., S\~ao Jos\'e dos Campos,  S\~ao Paulo, Brazil\\
S. C. Ying$^{\dag \ast}$\\
Dept. of Physics, Brown University, Providence, R.I.  U.S.A.  }

\date{\ }
\maketitle

\begin{abstract}
A two-dimensional \underline{Frenkel-Kontorova model} under a
steady external force is  used
to study the nonlinear sliding friction between flat macroscopic
surfaces  with
a  lubricant layer in between. The nonequilibrium properties of the
model
are simulated
by a Brownian molecular dynamics and results are obtained as a
function of
temperature and a microscopic friction parameter $\eta$.
\end{abstract}

\newpage

\section{Introduction}

The behavior of an adsorbed layer under an external force in the plane
is a
fundamental problem for understanding  the mechanisms responsible
for friction and lubrication between two flat macroscopic surfaces
\cite{Israelachvili,Persson}. Because
of the lubricant, the friction between the surfaces is essentially
determined by the force required to shear the adsorbed layer.
For small external force,
this is just the familiar study of the collective diffusion of the overlayer.
However, for strong external fields, even a pinned layer can acquire
significant mobility at low temperatures due to the lowering of the
diffusion barrier by the external field. Moreover, the steady state
effective temperature of the overlayer can be significantly higher than
that of
the  substrate.
Thus we can actually be dealing with the flow of a
liquid layer rather than simple diffusion in the solid phase.
Recently, there have been a great deal of interest in the study of
this problem in regards to boundary lubrication.
In particular,  Persson \cite{Persson} has made a series
of studies based on a model of adatoms  interacting with
Lennard-Jones pair
interaction. It is found  that the response of the overlayer can develop
hysteresis. This means that the threshold field $F_a$ required to
initiate
the motion of  the  adlayer is significantly larger than the threshold
force
$F_b$ below which an initially sliding layer will be pinned again. In fact,
based on general hydrodynamical considerations, it is argued  that the
ratio
of $F_a$ to $F_b$  should be 2 for all systems with sizable frictional
coupling to the substrate. This hypothesis was tested for the Lennard
Jones
system by direct numerical  simulation and seem to be
satisfied \cite{Persson}. The universality of
this relation between sliding friction and
static friction still remains an open question.
Even the existence of a hysteresis effect is not
necessarily true for other models of adsorbed layers.

We introduce a simple model for sliding friction based on a
two dimensional version of the Frenkel-Kontorova model
\cite{Frenkel},
where particles are coupled to  each other by elastic springs and to  a
periodic potential.
The equilibrium properties of this model are well understood both in
one and
two dimensions, largely in relation to the study of commensurate and
incommensurate transitions \cite{Pokrovsky}. It has also been a
standard model in
the study
of mobility of charge density waves \cite{Gruner}.
The infinite range of the harmonic interaction in the model leads to a
disordered (fluid) state very different from the
conventional  one  where atoms interact via short range forces. Even
in the
fluid phase, the atoms cannot diffuse around and will maintain their
original
neighbours. Thus it is of particular interest to test whether the
universality
of hysteresis and  the relation between static and kinetic friction still
applies to this simple model.

\section{Model and Simulation}

The simplest version of the Frenkel-Kontorova model
in two dimensions is to limit the displacements of the individual atoms
in one
direction only. Thus  the Hamiltonian is of the form

\begin{eqnarray}
H &=& \sum_{i,j} \left\{
{p_{ij}^{2}\over 2m}
+ {K_x\over 2} (x_{i+1,j} - x_{ij} - b)^2 +{K_y\over 2} (x_{i,j+1}-
x_{ij})^2 \right. \nonumber\\
&-& \left. U_0 \cos(2\pi x_{ij}/a)
 \right\}
\end{eqnarray}
which can be taken to represent a system of coupled linear harmonic
chains of
particles,  each of mass $m$ moving in a periodic potential of period
$a$ and
amplitude $U_o$. The second term represents the elastic interaction
within a
chain and the third term represents interaction between adatoms on
different
chains. The index ${i,j}$  labels the particle  with integer $i$
($ i=1,..., N_x $ ) and the chain with integer $j$ ($j=1,...,N_y$), so that
the system consists of $N_x \times N_y$ adsorbates at positions
$x_{i,j}$.
We have considered  $N_x=N_y= N$ and $K_x=K_y=K$.

Most of our results are  obtained assuming periodic boundary
conditions which
means that the ${N+1,j}$ particle is a periodic image of particle ${1,j}$
in each chain,  so
that $x_{N+1,j} = x_{1,j} + N_s \ a$, where $ N_s $ is the number of
local minima of the periodic substrate in the $x$-direction.
In the other direction, periodic boundary condition means that
$x_{i,N+1}$ = $x_{i,1}$.
The average distance between particles, $b$,  in each  chain
is determined through the boundary condition and is given by
$b = N_s \  a/N$ and the overlayer coverage can be
defined as  $\theta = N/N_s$. With this setup, $b$ and $\theta$ cannot
 be
changed independently and the ratio
$b/a$ is forced to be  rational.
We have also performed few calculations with free boundary
conditions in
the $x$-directions. In this case, the ratio b/a can assume arbitrary
nonrational values.               The system now can  remain
commensurate
for
a given $\theta$ only up to a critical value $\delta_c$ of the misfit
parameter \cite{Frenkel} $\delta = (b-a)/a$.

To study sliding friction in the model we assume that the substrate
acts as a
heat bath at an equilibrium temperature $T$ and that an external force
$F$
acts on each of the adsorbates. The equation of motion for the
particle coordinates are described by Brownian dynamics

\begin{equation}
m \ddot x_{ij} + m \eta \dot x_{ij} = -{\partial V \over \partial x_{ij} }
 -{\partial U \over \partial x_{ij}}  + F + f_{ij}
\end{equation}
where $U$ is the periodic potential, $V $ is the adsorbate-adsorbate
harmonic
interaction potential and $f_{ij}$ is a  random force that is related
to the microscopic friction parameter $\eta$ by the
fluctuation dissipation theorem
\begin{equation}
< f_{ij}(t) f_{i'j'}(t^\prime)> =  2 \eta m k_B T \
\delta_{i,i'} \delta_{j,j'} \delta \ (t-t^\prime)
\end{equation}

We have studied the sliding friction of the adsorbate by simulating the
above
equations using standard methods of Brownian molecular dynamics
\cite{Allen}.
We use units in which $a=1$, $m=1$ and $U_o=1$. Typically, the time
variable
was discretized with time step $\delta t = 0.02 \tau $ to $0.06 \tau$
where
$\tau = (m a^2/U_o)^{1/2}$ and $ 4 \times 10^5$ time steps were used in
each
calculation allowing $ 10^5$ time steps for equilibration. The elastic
constant was set to $K=10$ and
calculations were performed as function of temperature, external
force and
coverage for a system containing $N \times N$ adsorbate particles
where $N=10$.
The system was allowed to evolve into a steady state such that the
time average of physical quantities approached a constant.

\section{Results and Discussion}
When the external $F=0$, the  drift velocity $v_d$ of the adsorbate
center of
mass
is zero independent of the temperature with each particle performing
a
Brownian motion. At $T=0 K$, the velocity remains zero for any applied
force less than a critical value $F_o$ where the effective
potential minima disappears, $F_o = 2 \pi U_o$, the same result as for
the
one-dimensional case \cite{Pokrovsky}. From now on, we will normalize
the
external force by this critical value.
If a small force is applied at higher enough temperature,
in addition to the irregular motion of the particles, one expects  an
overall drift in the direction of the force with
a speed $v_d$ proportional to the external force, $v_d = F/m \bar \eta$
in
the linear regime, where $\bar \eta$ is an effective sliding friction. Fig.\
1  shows the results of the inverse of the normalized
linear sliding friction, $\eta/\bar \eta$, calculated as a function of
temperature for a small value of the external force, $F/F_o=0.035$ and
$b/a=2$. At low temperatures the adlayer is in a pinned state at small
forces, and $\eta/\bar \eta$ is essentially zero.        As the temperature
is increased $1/\bar \eta$
remains zero until $T=T_c$, beyond which $\eta/ \bar \eta$ increases
significantly and should approach $1$ at higher enough $T$. The value
of
$T_c$ correspond to the commensurate
solid-fluid    transition where
the overlayer at high temperatures become disordered and  is
effectively
depinned from the substrate. Figs.
2 and 3 shows snapshot pictures of the adsorbate
in the commensurate solid and the fluid      state respectively.
 From Fig.\ 1
we estimate that  $T_c \approx 2.3$. We have also performed similar
calculations with different boundary conditions.
 In Fig.\ 4 we show the results for the
structure factor at low and high temperatures for $b/a=1.9$ with free
boundary condition. At
low $T$ the structure factor has a sharp peak at $q=\pi$ and the
overlayer is
commensurate with the substrate with lattice vector
$2 a$. At high $T$ the structure factor is much broader corresponding
to a
fluid state. For $b/a=1.9$ it peaks at an incommensurate value which is
temperature dependent and limited by $q = 2 \pi/b$.   The
corresponding
result for $b/a=2$ and p.b.c is similar except that the structure factor
at high temperature remains peaked at  $q=\pi$.

In the nonlinear regime, the relation between $v_d$ and $F$ is
expected
to depend strongly the initial phase \cite{Persson}. If the adsorbate,
when
$F=0$, is a fluid, $v_d$ will be nonzero for arbitrarily
small external force. This result is also true for the present model.
In the fluid phase above $T_c$, the overlayer is essentially depinned
and any small
force leads to nonzero $v_d$. Moreover, the relation $v_d = f(F)$
shows no
hysteresis.
For $T < T_c$ however, where a commensurate (solid phase) prevails
when
$F=0$, a critical force $F_a$ is necessary to depin and initiate sliding.
It is expected \cite{Persson} that in this case the nonlinear sliding
friction
exhibit hysteresis as a function of $F$, i.e., the relation between
velocity and
external force depends on  whether $F$ increases from zero or
decreases from
a high value and this  in turn  implies a stick and slip motion.

In Fig.\ 5, we show the results for $v_d $ as a function of $F$ for
$\eta =0.6$, $T=0.5$. Note that in this case the temperature
is considerably less than the  activation energy barrier ($2U_0$) for
particle
diffusion. We found that this is an important requirement  for
hysteresis to appear. Initially, the adlayer is in a pinned state.
Then the simulation was started with  various external force
strengths.
An hysteresis loop is clearly seen, corresponding to a stick and slip
motion. The effective temperature $T^*$ of the adlayer is shown
Fig.\ 6. It is  seen
that the initial sliding phase when the applied force is increased
beyond the static threshold $F_a$ has almost the same temperature
as the
substrate and the substrate potential provides no additional
resistance.
This is a sliding solid phase. It corresponds to the floating or
incommensurate
solid phase in the equilibrium situation. It is important to note that this
phase is dynamically generated by the external
field. With our present choice of
parameters, the misfit is zero and the floating solid phase does not
exist in
equilibrium.
Thus the external force induces a commensurate solid to sliding
solid transition at $F=F_a$. As the force is decreased through $F_a$,
there
is a jump in $T^*$ accompanied by a simultaneous decrease in $v_d$.
This corresponds to an overheating of overlayer to an effective
temperature
above $T_c$ (compare with Fig.\ 1), dynamically ``melting"  the
sliding solid phase. This can be regarded as sliding solid to  fluid
transition. Finally at
$F=F_b$, the fluid phase condenses into a commensurate solid phase
and the
temperature of the adlayer and the substrate become the same again.
The fact that there is a condensation from the "melted" phase
is confirmed by noting  that the temperature of the adlayer at this
condensation is almost identical to the
temperature $T_c$, the temperature where the  pinned  solid to fluid
phase
transition occurs. Thus  there is already a rich variety of dynamical
phase
transitions manifested in this simple model. We note that Persson
\cite{Persson}
has observed very similar phase transitions for Lennard-Jones
systems,
showing that these features are universal in the non-linear response
of
overlayers and  not just for specific models. In addition the present
results show that a conventional  "liquid"  phase is not required and a
high
temperature  disordered fluid
phase with vanishing shear resistance  is sufficient to show some of
these effects.

We now consider the behavior of the nonlinear sliding friction at
higher
temperatures but still less than $T_c$. We find that at $T \approx U_o$,
hysteresis disappears  as indicated in Fig.\ 7 but
$F_a$ can still be defined. This behavior
is not specific   to      the present model and has also been found in
the
molecular dynamics simulation of the Lennard-Jones system
\cite{Baldan}.
This temperature is comparable to the energy barrier for center of the
mass
diffusion and for a given force $F$, within the range where hysteresis
is
observed at low $T$, the system is able to relax in a short enough time
to a stable single  velocity, independent of the history of force
variation.
This suggests that the hysteresis observed at low $T$ is due to
metastability.

We have also studied the effects of varying the microscopic friction
parameter $\eta$ and found that the  ratio $F_b/F_a$  strongly
depends on
$\eta$ and converge to $F_b/F_a=1$ for sufficiently large $\eta$ as
shown in Fig.\ 8. This
should be contrasted to the results obtained by Persson
\cite{Persson}
for the Lennard-Jones system where $F_b/F_a$ is found to saturate
for
large $\eta$ at a universal value $\approx 0.6$.
We suggest that this is due to the fact that the
nature of the fluid phase in the two  models are very differrent. The
hydrodynamics    arguments
used by Persson \cite{Persson} based on the existence of a drag force
acting
on a nucleating domain may not be valid in this case.

\section{Conclusions}

There are many questions that remains to be investigated. The  most
important
one is probably to establish which features of the nonlinear transport
are
universal and model independent and which ones are model sensitive.
Already, we
have observed  in the  Frenkel-Kontorova model that under increasing
values of the microscopic
friction $\eta$, the hysteresis loop tends to disappear  whereas it is
stable in
the Lennard Jones system. We suspect that this is related  to the
stability of
the sliding solid phase as a function of $\eta$ as well as the
dependence of the
fluid-pinned solid transition  on the nature of the fluid state.
However, many features observed in the                Lennard-Jones
system
also appear in the simple Frenkel-Kontorova model.
The advantage of this  model  system is that the equilibrium property
of the
commensurate-incommensurate transition  is well understood
through a series of studies.\hfil\break

$\dag$ Research supported in part by a joint NSF-CNPq grant.

$\ast$ Research supported in part by an ONR grant.

\newpage
\begin{figure}
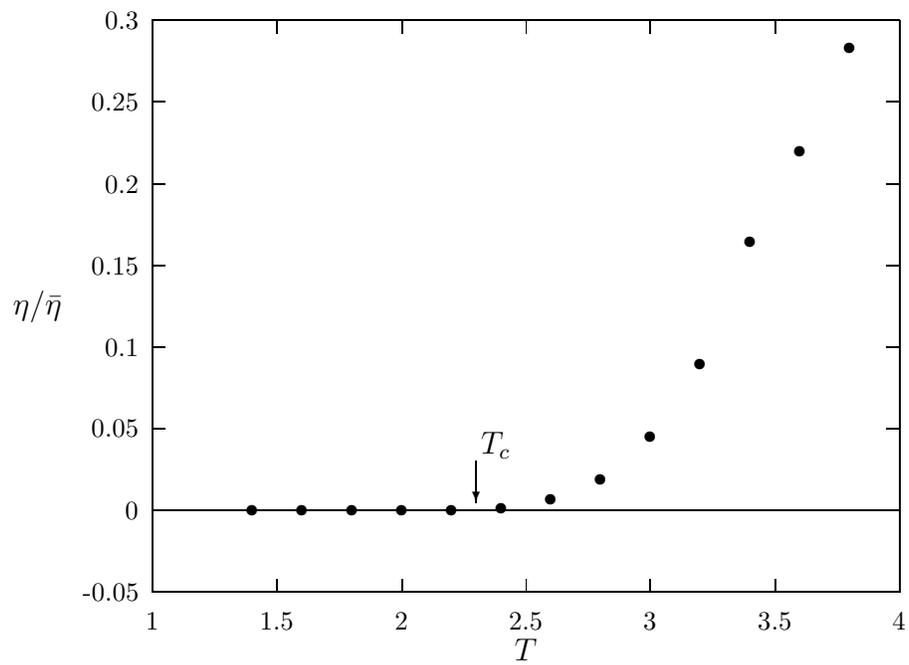

\begin{center}
\setlength{\unitlength}{0.240900pt}
\ifx\plotpoint\undefined\newsavebox{\plotpoint}\fi
\sbox{\plotpoint}{\rule[-0.175pt]{0.350pt}{0.350pt}}%


\end{center}
\caption{Inverse of the normalized sliding friction $\bar \eta$ as a function
of temperature $T$ for $\eta = 0.5$, $b/a=2$}.
\label{vxt}
\end{figure}

\begin{figure}
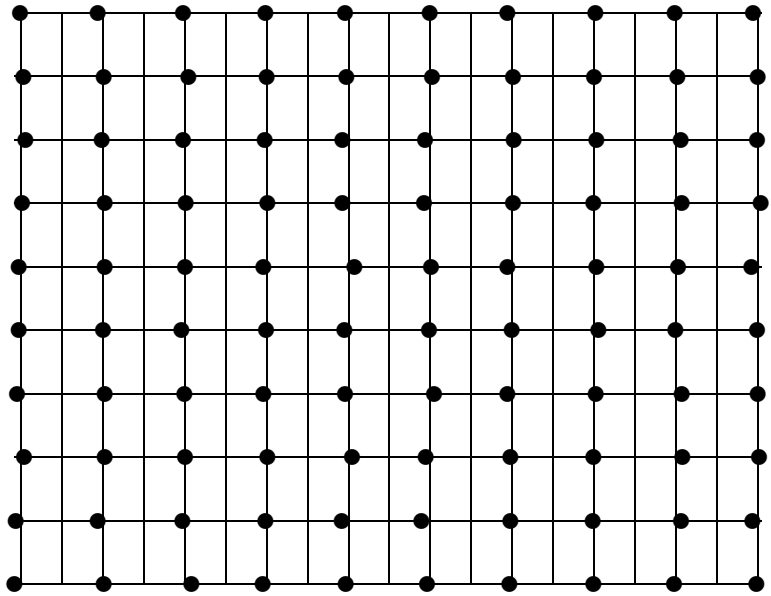

\begin{center}
\setlength{\unitlength}{0.240900pt}
\ifx\plotpoint\undefined\newsavebox{\plotpoint}\fi
\sbox{\plotpoint}{\rule[-0.175pt]{0.350pt}{0.350pt}}%
%
\end{center}
\caption{Snapshot picture of the adsorbate in the commensurate solid phase,
for $T=0.5$, $b/a=2$.}
\label{configT05}
\end{figure}

\begin{figure}
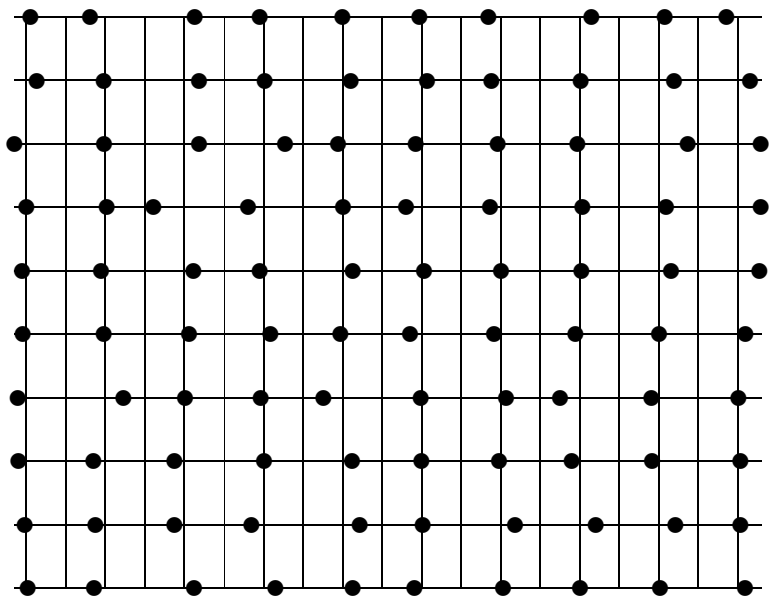

\begin{center}
\setlength{\unitlength}{0.240900pt}
\ifx\plotpoint\undefined\newsavebox{\plotpoint}\fi
\sbox{\plotpoint}{\rule[-0.175pt]{0.350pt}{0.350pt}}%
%

\end{center}
\caption{Snapshot picture of the adsorbate in the high temperature phase,
for $T=3$, $b/a=2$.}
\label{configT3}
\end{figure}

\begin{figure}
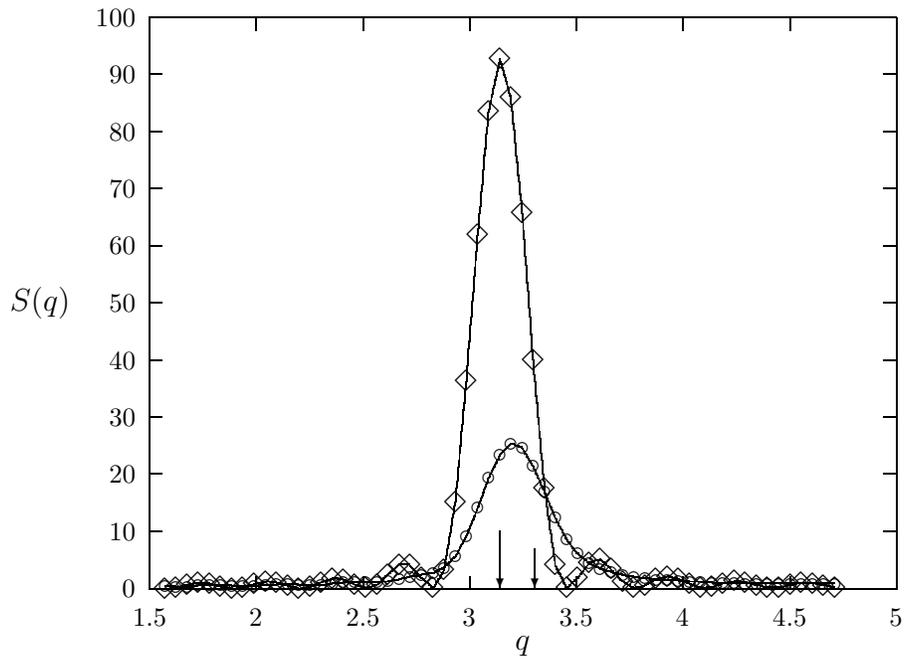

\begin{center}
\setlength{\unitlength}{0.240900pt}
\ifx\plotpoint\undefined\newsavebox{\plotpoint}\fi
\sbox{\plotpoint}{\rule[-0.175pt]{0.350pt}{0.350pt}}%
%

\end{center}
\caption{Structure factor $S(q)$ as a function of reciprocal lattice
vector $q$ at different temperatures, using free boundary conditions
and $b/a = 1.9$. Open  diamonds  correspond to $T=0.5$
and  circles to $T=3$. The large and small arrows indicate reciprocal
lattice vectors $q = 2 \pi/2 a$ and $q = 2 \pi/b$,  respectively.}
\label{struc}
\end{figure}

\begin{figure}
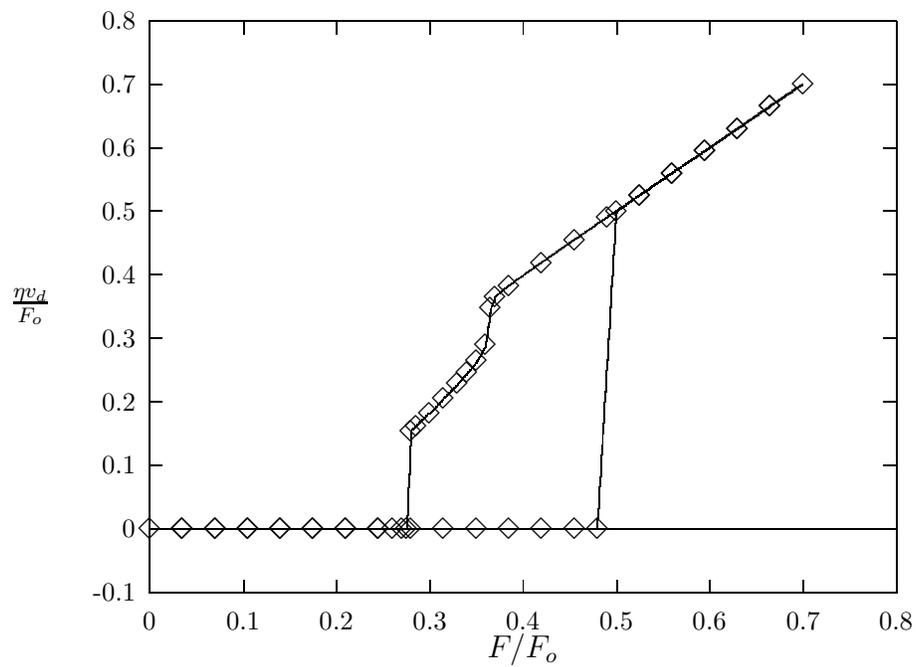

\begin{center}
\setlength{\unitlength}{0.240900pt}
\ifx\plotpoint\undefined\newsavebox{\plotpoint}\fi
\sbox{\plotpoint}{\rule[-0.175pt]{0.350pt}{0.350pt}}%
%

\end{center}
\caption{Drif velocity $v_d$ as a function of external force $F$ for
$\eta = 0.6$ and $T=0.5$.}
\label{vxfg06T05}
\end{figure}

\begin{figure}
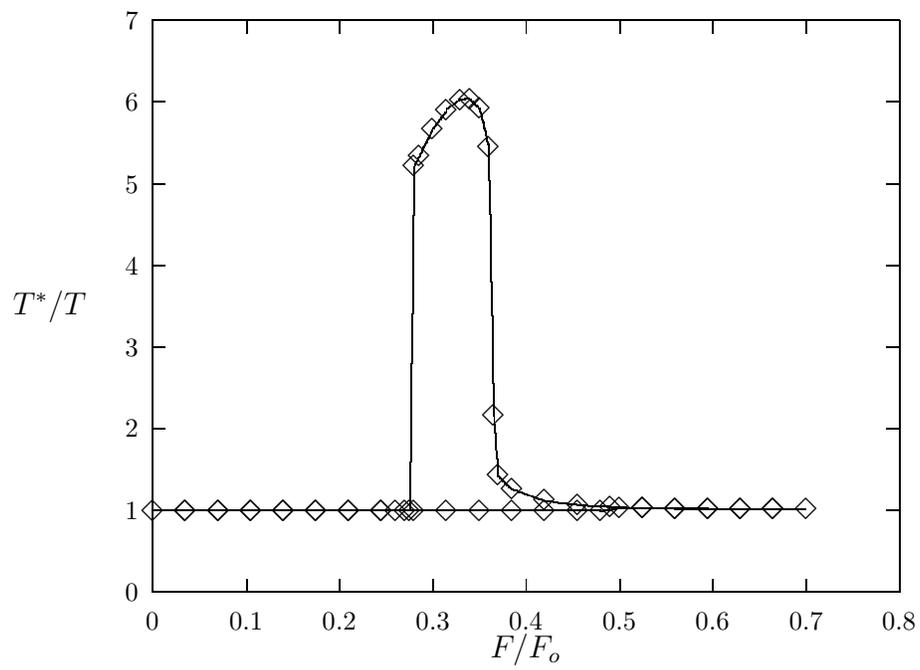

\begin{center}
\setlength{\unitlength}{0.240900pt}
\ifx\plotpoint\undefined\newsavebox{\plotpoint}\fi
\sbox{\plotpoint}{\rule[-0.175pt]{0.350pt}{0.350pt}}%
%

\end{center}
\caption{Effective temperature $T^*$ of the overlayer normalized to
the substrate temperature $T$ as a function
of the external force $F$ for $\eta=0.6$ and $T=0.5$.}
\label{teffg06T05}
\end{figure}

\begin{figure}
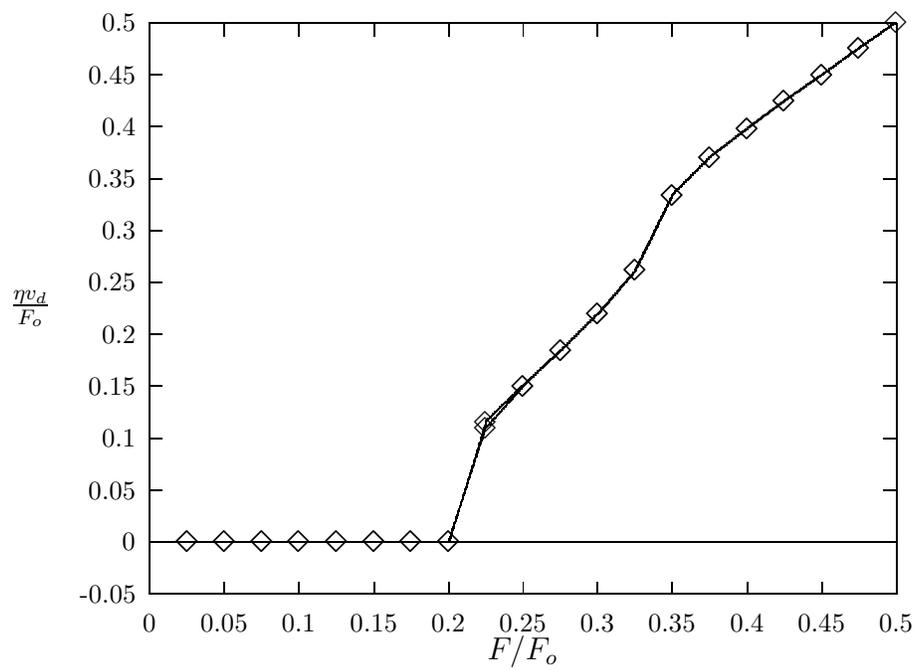

\begin{center}
\setlength{\unitlength}{0.240900pt}
\ifx\plotpoint\undefined\newsavebox{\plotpoint}\fi
\sbox{\plotpoint}{\rule[-0.175pt]{0.350pt}{0.350pt}}%
%

\end{center}
\caption{Drif velocity $v_d$ as a function of external force $F$ for
$T=1$ and $\eta = 0.5$.}
\label{vxfg05T1}
\end{figure}

\begin{figure}
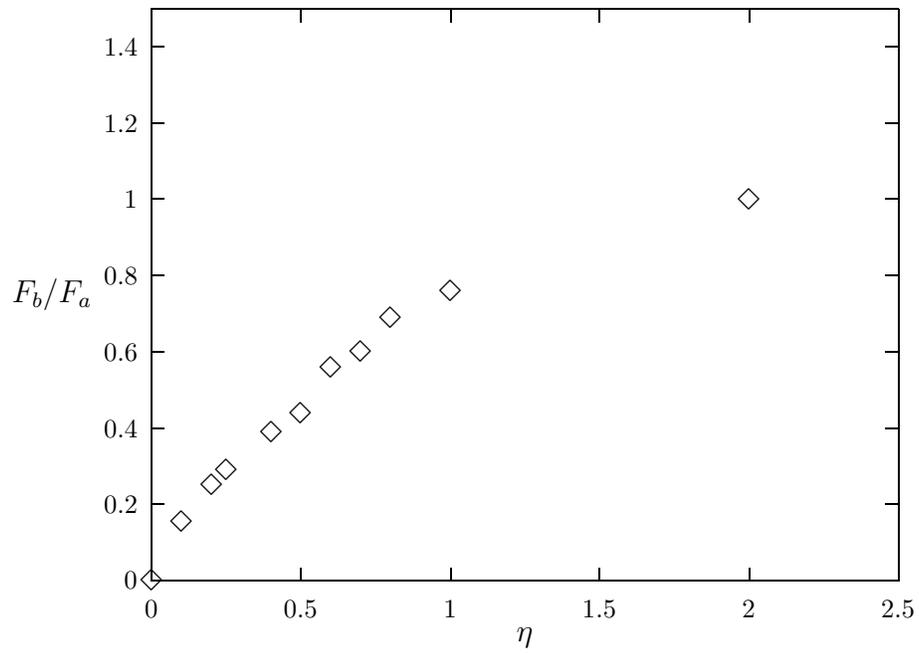

\begin{center}
\setlength{\unitlength}{0.240900pt}
\ifx\plotpoint\undefined\newsavebox{\plotpoint}\fi
\sbox{\plotpoint}{\rule[-0.175pt]{0.350pt}{0.350pt}}%
%
\caption{Ratio $F_b/F_a$ as a function of $\eta$ for $T=0.5$}
\end{center}
\label{ratio}
\end{figure}

\end{document}